\documentclass[twocolumn]{aastex631}

\usepackage[version=4]{mhchem}
\usepackage{tikz}

\newcommand{\cs}{c_{\rm s}}
\newcommand{\degree}{^{\circ}}

\newcommand{\kep}{\text{Keplerian}}
\newcommand{\kms}{\text{km/s}}

\newcommand{\Msun}{M_\odot}
\newcommand{\Mstar}{M_\star}

\newcommand{\Mj}{M_\mathrm{J}}
\newcommand{\Mth}{M_\mathrm{th}} 
\newcommand{\nphot}{N_{\rm{photon}}}
\newcommand{\panel}[1]{\text{panel #1}} 
\newcommand{\phip}{\phi_{\mathrm{p}}} 
\newcommand{\rhill}{R_{\mathrm{H}}} 

\newcommand{\rp}{r_{\mathrm{p}}} 
\newcommand{\Rstar}{R_\star}
\newcommand{\Rsun}{R_\odot}
\newcommand{\Tstar}{T_\star}
\newcommand{\vch}{V_{\rm{ch}}}

\newcommand{\vkep}{V_{\rm{Kep}}}

\newcommand{\vkepp}{V_{\rm{p,Kep}}}     
\newcommand{\vphi}{V_{\phi}}
\newcommand{\vpd}{V_{\phi}-<V_{\phi}>} 
\newcommand{\vrr}{V_{r}}
\newcommand{\vrd}{V_{r}-<V_{r}>}   
\newcommand{\vtheta}{V_{\theta}}
\newcommand{\vtd}{V_{\theta}-<V_{\theta}>} 

\newcommand{\vlosp}{V_{\rm{p,LoS}}}

\newcommand{\fargo}{\text{FARGO3D}}

\newcommand{\radmc}{\text{RADMC3D}}

\begin{document}

\title{Mind the kinematics simulation of planet-disk interactions: 

time evolution and numerical resolution}

\author[0000-0002-8095-7448]{Kan Chen}
\affiliation{Department of Physics and Astronomy, University College London, Gower Street, London, WC1E 6BT, UK, kan.chen.21@ucl.ac.uk}

\author[0000-0001-9290-7846]{Ruobing Dong}
\affiliation{Kavli Institute for Astronomy and Astrophysics, Peking University, Beijing 100871, People’s Republic of China, rbdong@pku.edu.cn}
\affiliation{Department of Physics and Astronomy, University of Victoria, Victoria, BC, V8P 5C2, Canada}

\begin{abstract}

Planet-disk interactions can produce kinematic signatures in protoplanetary disks.
While recent observations have detected non-$\kep$ gas motions in disks,
their origins
are still being debated. To explore this, we conduct 3D hydrodynamic simulations using the code $\fargo$ to study non-axisymmetric kinematic perturbations at 2 scale heights induced by Jovian planets in protoplanetary disks, followed by examinations of detectable signals in synthetic CO emission line observations at millimeter wavelengths. We advocate for using residual velocity or channel maps, generated by subtracting an azimuthally averaged background of the disk, to identify planet-induced kinematic perturbations.
We investigate the effects of two basic simulation parameters, simulation duration and numerical resolution, on the simulation results. 
Our findings suggest that a short simulation (e.g., 100 orbits) is insufficient to establish a steady velocity pattern given our chosen viscosity ($\alpha=10^{-3}$), and displays plenty of fluctuations on orbital timescale. 
Such transient features could be detected in observations.
By contrast, a long simulation (e.g., 1,000 orbits) is required to reach steady state in kinematic structures. At 1,000 orbits, the strongest and detectable velocity structures are found in the spiral wakes close to the planet.
Through numerical convergence tests,
we find hydrodynamics results 
converge in spiral regions at a resolution of 14 cells per disk scale height (CPH) or higher. Meanwhile, synthetic observations produced from hydrodynamic simulations at different resolutions are indistinguishable with 0.1$\arcsec$ angular resolution and 10 hours of integration time on ALMA. 

\end{abstract}


\section{Introduction}

When planets form in protoplanetary disks, they can perturb the gas motion \citep{goodman_planetary_2001}, and produce kinematic signatures in line emission observations \citep{perez_planet_2015, perez_observability_2018}.
By searching for and characterising such signatures, forming planets in disks can be identified, and their properties, such as masses and locations, can be constrained \citep{pinte_kinematic_2018, casassus_kinematic_2019, teague_spiral_2019}. 
This mechanism has become increasingly important in the search for planets in disks, which are challenging to find using more conventional planet detection methods such as radial velocity and transit surveys.

Among the various types of planet-induced kinematic signatures, a zigzag structure in the isovelocity curve in channel maps of gas emission, a.k.a. a ``kink'', has resulted in the most planet detections. Over a dozen planet candidates have been found in this way \citep{pinte_kinematic_2018, pinte_kinematic_2019, pinte_nine_2020}. \citet{bollati_theory_2021} have developed analytical models to quantify expected signals, while \citet{izquierdo_disc_2021} have proposed a statistical framework to quantify their detections in observations. Some of the planets discovered through the kink signature have been incorporated into the NASA’s Exoplanet Database (e.g., HD 97048 b from \citet{pinte_kinematic_2019}). Related observational signatures and planet detection techniques, such as ``Doppler flip'', have been developed and successfully applied to real systems as well \citep{casassus_kinematic_2019}.

Despite the successes, a number of issues exist and prevent establishing more robust and definitive connections between observed kinematic signals and planets, as highlighted in the recent review by \citet{pinte_kinematic_2022}. One of the most important questions is the exact origin of the ``kink'', which has not been fully determined. 
Also, it is unclear what the best strategy is to quantify the statistical significance of detected kink signals. Finally, in some systems the expected spirals in millimeter dust emission associated with the kink planets were not detected \citep{speedie_testing_2022}, raising questions about our understanding of the observed kinematic signatures. 

To address these issues, numerical simulations of disk-planet interactions are needed to reproduce and analyze planet-induced kinematic signatures \citep{disk_dynamics_collaboration_visualizing_2020}. This is necessary as most planets inferred from their kinematic signatures are above the disk thermal mass \citep{speedie_testing_2022}, a regime where quantitative analytical theories on disk-planet interactions lack.
A number of works have investigated planet-induced kinematic signatures in simulations. \citet{perez_planet_2015} and \citet{ pinte_kinematic_2018, pinte_kinematic_2019, pinte_kinematic_2023} employed smooth particle hydrodynamic (SPH) simulations; with the exception of \citet{pinte_kinematic_2019} these simulations are up to 100 planetary orbits. \citet{perez_observability_2018} used a grid-based code, $\fargo$, and focused on velocity perturbations at the midplane and their observational signatures. \citet{rabago_constraining_2021} also used a grid-based hydro code (Athena$++$), coupled with a high $\alpha$ viscosity (0.01). They focused on velocity perturbations in hydro calculations at different vertical layers instead of observational signatures. 

Previous simulations provide insights into the kinematic signals induced by planets. However, the effects of two basic parameters, numerical resolution and system evolution time,
have not been thoroughly investigated. The goal of this work is to study how planet-induced signatures in simulations depend on the two parameters.
Meanwhile, we focus on non-axisymmetric features because they are expected to be more helpful in locating embedded planets than axisymmetric features, and because most reported planet detections so far are based on such features.

The paper is organized as follows. 
In \S\ref{sec:hd}, we lay out the setup for 3D hydro simulations. \S\ref{sec: hydro results} presents the results of planet-induced velocity perturbations. We study the time evolution of simulation and carry out numerical resolution convergence tests. In \S\ref{sec:rt}, we make synthetic observations to further explore how the simulation time and numerical resolution can affect the planet-induced kinematics signatures. We summarize the results in \S\ref{sec: conclusion}.

\section{Hydrodynamic simulation setup}
\label{sec:hd}

We conduct 3D gas-only hydrodynamic (HD) simulations in spherical coordinates ($r$, $\phi$, $\theta)$ = (radial, azimuthal, colatitude) using the grid-based code $\fargo$ \citep{benitez-llambay_fargo3d_2016}.
A planet is fixed at a circular orbit at $\rp$. The simulation domain spans from $0.36 \rp$ to $2.75 \rp$ in $r$, $0$ to $2\pi$ in $\phi$, and $0.5\pi-0.35$ to $0.5\pi$ in $\theta$ (upper half-disk). Mesh grids are linearly distributed in $\phi$ and $\theta$, and logarithmically distributed in $r$.

For boundary conditions,
density and azimuthal velocities are extrapolated at the radial boundaries, while reflecting boundary conditions are applied in the colatitude direction to prevent mass inflow or outflow.
Damping zones are applied near the radial boundaries of the mesh \citep{de_val-borro_comparative_2006}. 

Following the approach of \citet{perez_observability_2018}, we assume an initial radial gas surface density profile of $\Sigma_g = \Sigma_{0} (r/\rp)^{-1}$, where $\rp=100$ au and $\Sigma_0 = 0.09$ g/cm$^{-2}$, resulting in a total disk mass of $1.5\times 10^{-3} \Msun$ within the simulation domain. We model a flared disk with an aspect ratio $h/r$ of 0.08 at $\rp$ and a flaring index of 0.15. The equation of state is assumed to be isothermal. We adopt an $\alpha$ viscosity of $10^{-3}$, consistent with observations \citep{flaherty_weak_2015,flaherty_three-dimensional_2017, flaherty_turbulence_2018}. The planet mass is chosen as 5$\times$ the disk thermal mass $\Mth=(h/ r)_{\mathrm{p}}^{3} M_{\star}$ (\citealt{rafikov_nonlinear_2002}), which is $\sim2.5\Mj$ around a solar-mass star given our chosen $(h/r)_{\mathrm{p}}$.
The planet mass is typical among those in real systems with kinematic signature-based planet detections \citep[Table A1 in][]{speedie_testing_2022}. The simulations include the indirect term in the stellar gravity.

Four physical quantities are calculated in $\fargo$ simulations: density ($\rho$), radial velocity ($\vrr$), azimuthal velocity ($\vphi$), and colatitude velocity ($\vtheta$). In visualization, we define the positive directions in velocities as moving away from the star for $\vrr$, counter-clockwise rotation for $\vphi$ (same as the Keplerian flow), and toward the midplane for $\vtheta$.

\section{Planet-induced perturbations in velocities} 
\label{sec: hydro results} 

We run the simulations for 1,000 orbits and study how the non-axisymmetric velocity perturbations induced by a planet evolve with time in \S\ref{sec:evolution}. We then carry out convergence tests to examine how the perturbations depend on numerical resolution in \S\ref{sec:convergence}. Our cells are cubic and we produce simulations with an effective resolution of 7, 10, 14, or 20 grid cells per scale height (CPH) in the region around the planet. As a reference, the simulation with a resolution of CPH$=20$, the default setting in \S\ref{sec:evolution}, has a grid of 500 $\times$ 1500 $\times$ 90 in ($r$, $\phi$, $\theta$). 

Most planet-induced kinematic signatures have been found in observations of CO line emission (Pinte et al. 2022), usually optically thick and from an emission surface at optical depth $\tau=1$. We locate the $\tau=1$ surface of CO $J=2-1$ emission at $\sim2h$ (Fig.~\ref{fig: tau1}; Appendix \ref{sec: tau1}), and focus on velocity perturbations at this surface.

\begin{figure}
\includegraphics[width=\columnwidth]{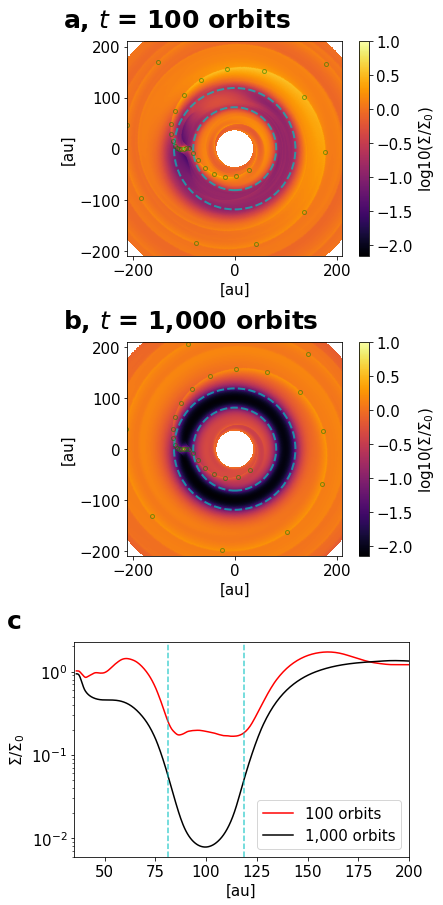} 
\caption{
Surface density at 100 (upper) and 1,000 orbits (middle), and the corresponding azimuthal average (excluding azimuth within $\arcsin \left(3 \rhill / \rp\right)$ to the planet) surface density profiles (lower). Cyan lines mark gap edges of $\rp \pm 2\rhill$ ($\rp=100$ au and $\rhill \sim 9.4$ au) in all panels. The surface density peak along the primary inner and outer spirals is traced out by green open circles.}
\label{fig: sigma_gap}
\end{figure}

Fig.~\ref{fig: sigma_gap} shows the surface density of the simulation with CPH$=20$ at 100 and 1,000 orbits. The planet launches spiral density waves in both the inner and outer disks, and gradually opens a gap. We note that although a super-thermal mass planet can excite multiple prominent spiral arms, in particular in the inner disk \citep{fung_inferring_2015, bae_planet-driven_2018, bae_planet-driven_2018-1}, we focus on the primary arm as it dominates the velocity perturbations in regions close to the planet. The gap approaches its asymptotic depth at $\sim1,000$ orbits at the given viscosity \citep{fung_gap_2016}. At earlier stages, e.g., 100 orbits, accumulation of gas at the Lagrange points L4 and L5 are visible, with the latter being more prominent.

\subsection{Temporal variations in planet-induced velocity perturbations} 
\label{sec:evolution} 

In our simulations, we have observed variabilities on both short (orbital) and long timescales (across 1,000 orbits) in planet-induced velocity perturbations. We separately discuss them in \S\ref{sec: fargo_time_short} and \S\ref{sec: fargo_time_long}.

\subsubsection{Local variabilities in $\vrr$ on orbital timescale}
\label{sec: fargo_time_short}

\begin{figure*}
\includegraphics[width=\linewidth]{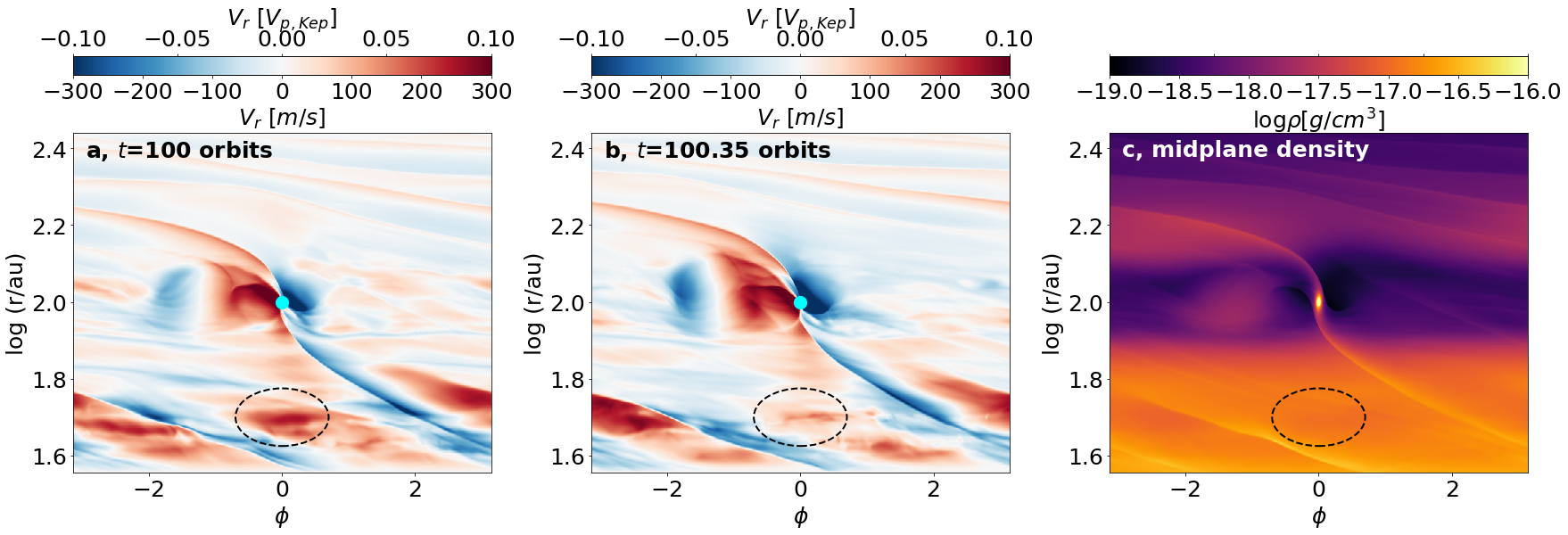} 
\centering
\caption{
$\vrr$ at $2h$ from the midplane in a simulation with a planet mass of 5$\Mth$ and a numerical resolution of CPH$=20$ at 100 (left) and 100.35 (middle) planetary orbits. The right panel shows the midplane density. The planet location is marked with a cyan dot in the first two panels. The dashed circle marks a region in between primary and secondary spiral arms in the inner disk. Colorbars are in units of m/s and planetary $\kep$ velocity $\vkepp$.
See \S\ref{sec: fargo_time_short} and \S\ref{sec: fargo_time_long} for discussions. 
} 
\label{fig: hd_100ob}
\end{figure*}

We focus on $\vrr$ in this part, a quantity close to zero everywhere in a smooth disk if the planet is not present. Figure~\ref{fig: hd_100ob} shows $\vrr$ in the $r-\phi$ plane at $2h$ (i.e., $\theta=\pi/2-2(h/r)_{\rm p}$) at 100 orbits (left) and 100.35 orbits (middle). The midplane density panel (c) shows the locations of the gap and spirals. 100 orbits has been the chosen epoch in previous studies of planet-induced kinematic signatures using SPH simulations \citep[e.g.,][]{pinte_kinematic_2018, terry_locating_2022}.
The dashed circle highlights a region in the inner disk in between the primary and secondary spirals.
At 100 orbits, this region exhibits $|\vrr|$ = 200 m/s = 0.07 $\vkepp$ ($\vkepp$ is the planetary Keplerian velocity), higher than those along the spirals (about 150 m/s). However, at 100.35 orbits, $|\vrr|$ in this region falls below 100 m/s (0.03 $\vkepp$), indicating that velocity perturbations have not reached a steady state. In addition, this shows that planet-induced spiral density waves do not always dominate in $\vrr$ perturbations at this time.

\begin{figure*}
\includegraphics[width=\linewidth]{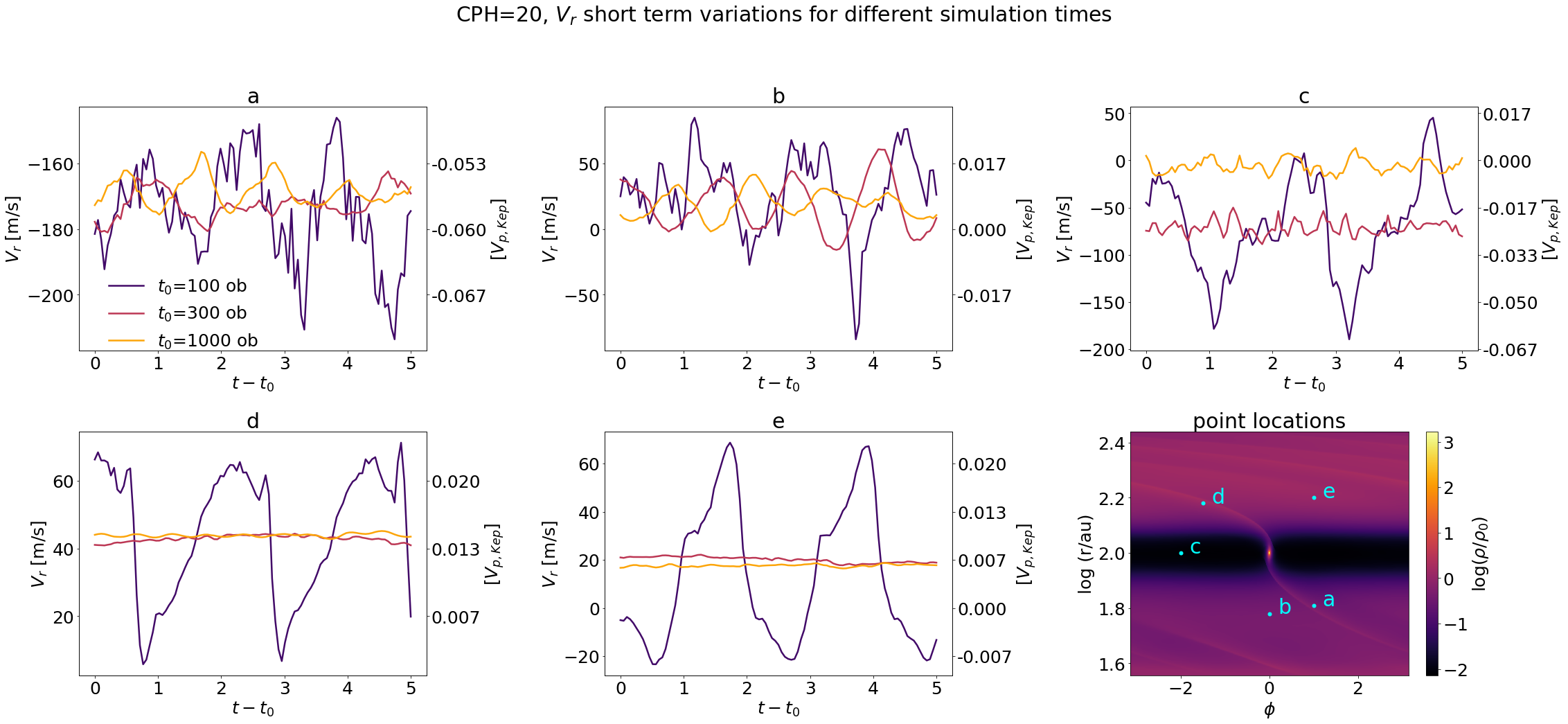} 
\centering
\caption{
Short term variations in $\vrr$ within 5 orbits at 5 locations at $2h$ above the midplane from point a to e in the disk and at $t_0=$100, 300, and 1,000 orbits. The locations of points a, b, c, d, and e are shown in the midplane density map at 1,000 orbits in the lower right panel.
See \S\ref{sec: fargo_time_short} for details. 
} 
\label{fig: vr_times}
\end{figure*}

The short term variability illustrated in Fig.~\ref{fig: hd_100ob} varies over the timescale of 1,000 orbits as well.
Fig.~\ref{fig: vr_times} shows $\vrr$ at $2h$ (solid lines) as a function of time within 5 orbits at 5 representative points a, b, c, d, and e in the disk at $t=100$, 300, and 1,000 orbits.
These points are located at an inner spiral (a), in between the primary and secondary spirals in the inner disk (b), inside the gap (c), at an outer spiral (d), and in between the spirals in the outer disk (e).
The temporal standard deviations (STD) at 100, 300, and 1,000 orbits are presented in Table \ref{tab: std}. We use outputs every 0.05 orbits in the plot and STD calculations. A straight line from panel (a) to panel (e) and an STD close to 0 indicate the flow is steady on the orbital timescale at the epoch. 

Overall, at 100 orbits, $\vrr$ exhibits significant short term variabilities, which tends to damp at 300 and 1,000 orbits. For example, at point b (in between spirals in the inner disk), $\vrr$ varies between -80 and 70 m/s within 5 orbits at 100 orbits (0.03 $\vkepp$), with an STD of 30 m/s (0.01 $\vkepp$). In contrast, $\vrr$ varies between 0 and $\sim50$ m/s at 300 orbits, with an
STD of 19 m/s (about 0.01 $\vkepp$), which drops further to 9 m/s at 1,000 orbits. 

\begin{figure*}
\includegraphics[width=\linewidth]{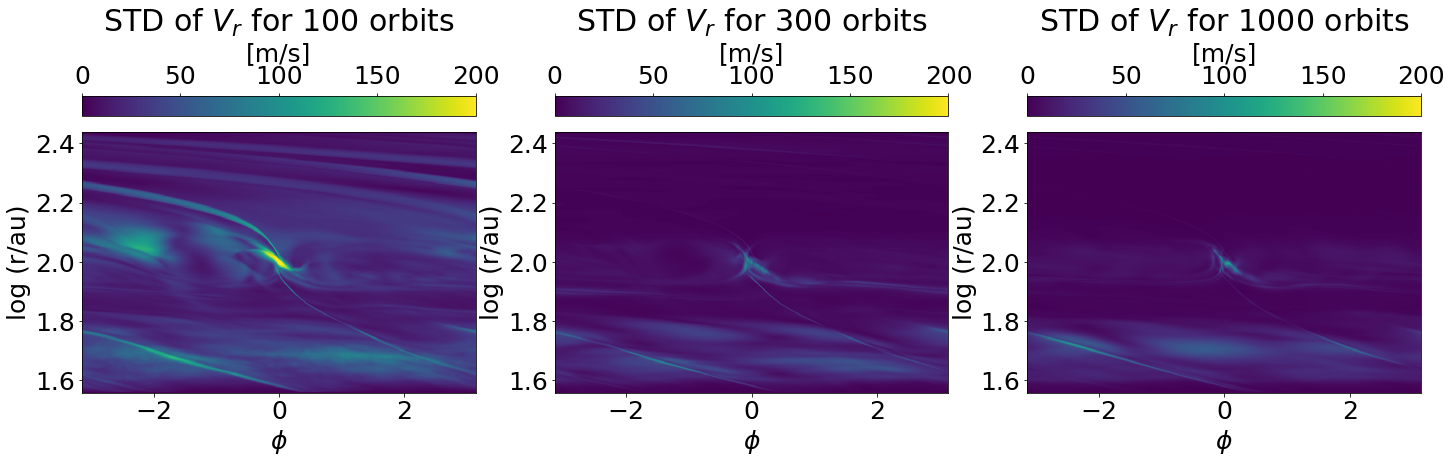} 
\centering
\caption{2D standard deviation map of $\vrr$ at 100, 300 and 1,000 orbits (left to right). The standard deviations are obtained within 5 orbits (see details in \S\ref{sec: fargo_time_short}).
}
\label{fig: std_2d} 
\end{figure*}

We also provide a 2D temporal STD map of $\vrr$ at 100, 300 and 1,000 orbits in Figure \ref{fig: std_2d} to show the variability in other regions. Overall, we can see the fluctuation damps with time. At 100 orbits, regions near the planet, the gap, and the spiral arms exhibit significant variability, with STDs topping 100 m/s. However, as the system evolves over 300 and 1,000 orbits, the $\vrr$ field becomes nearly steady across most regions.

In addition, we show the short term variability of $\vphi - \vkep$ and $\vtheta$ in Appendix \ref{sec: short_term_others}. The trend in both quantities is similar to that observed in $\vrr$.

\begin{table}[]
\caption{Temporal standard deviations (STD) of $\vrr$ shown in Fig.~\ref{fig: vr_times} ($\vrr$ within 5 orbits at 5 different locations at $t = 100$, 300, and 1,000 orbits).}
\label{tab: std}
\begin{tabular}{llllll}
\hline
\hline
STD (m/s)    & point a  & b & c  & d  & e  \\ \hline
t=100 (orbits)  & 15 & 30 & 51 & 18 & 28 \\ \hline
t=300  & 4  & 19 & 7  & $<1$  & 1  \\ \hline
t=1,000 & 4  & 9  & 7  & $<1$  & $<1$  \\ \hline
\end{tabular}
\end{table}

\subsubsection{Global signatures and their long term trend}
\label{sec: fargo_time_long}

\begin{figure*}
\includegraphics[width=\linewidth]{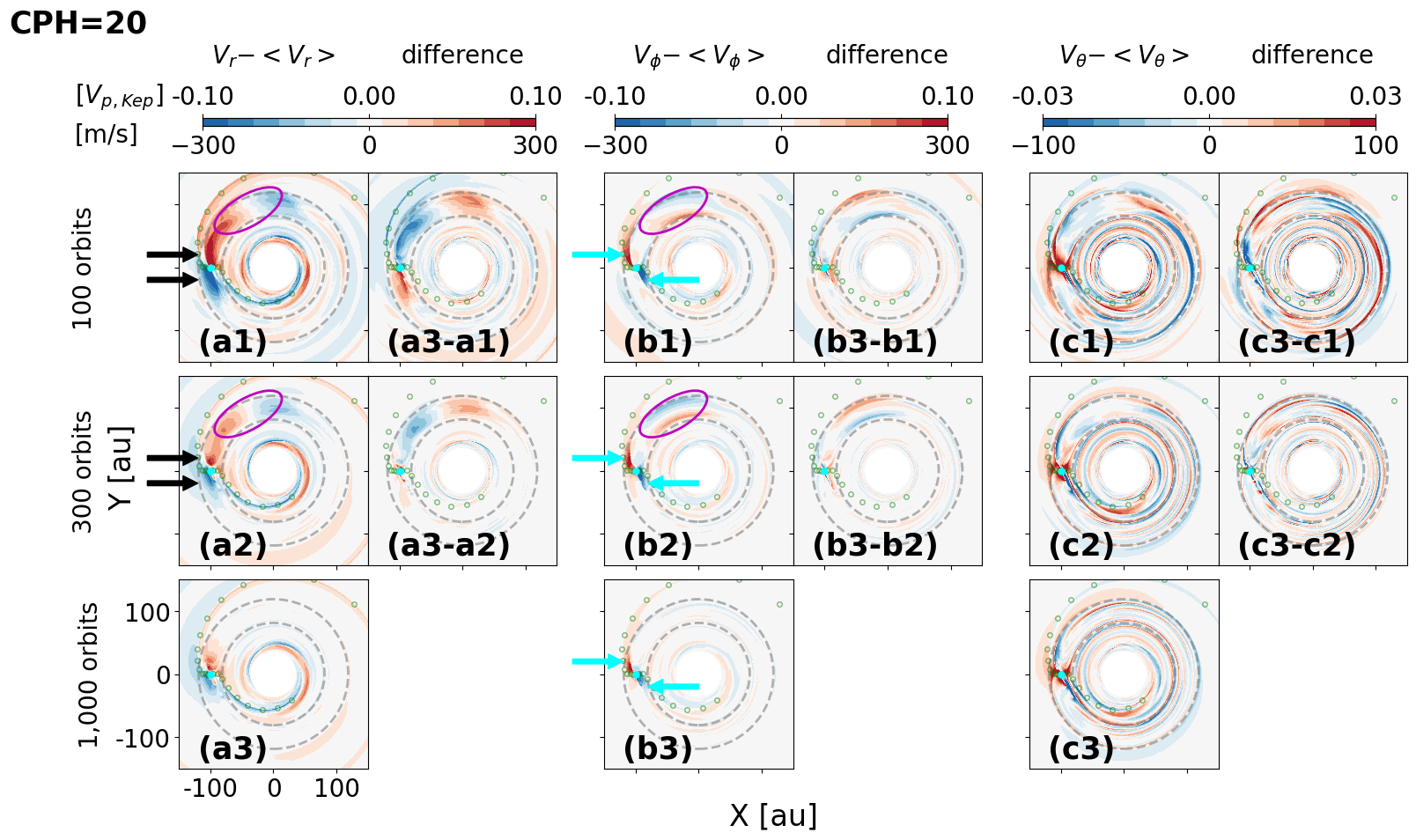} 
\centering
\caption{
Group (a) panels: The non-axisymmetric components in $\vrr$ at $2h$ at 100, 300, and 1,000 orbits (left column), and their differences (right column). Group (b) and (c) panels: similar to group (a) panels, but for $\vphi$ (b) and $\vtheta$ (c).
The location of the planet is indicated by the cyan marker, and its size is set to 0.4$\rhill$, the expected size of the circumplanetary disk. 
The trajectories of the spirals in the surface density are indicated by the small green circles ((also shown in Fig.~\ref{fig: sigma_gap}). Gap edges at $r=\rp-2\rhill$ and $\rp+2\rhill$ are marked with grey dashed circles. The magenta ellipse indicates the gas structure at L5. The two ends of the horseshoe flow inside the gap are indicated by black arrows and the strong converging flows are indicated by cyan arrows. The colorbars are in linear scale. See \S\ref{sec: fargo_time_long} for details.}
\label{fig: fargo_time}
\end{figure*}

We examine the perturbations in all three velocities at $2h$ from the midplane at 100, 300 and 1,000 orbits in Fig. \ref{fig: fargo_time}. To eliminate the strong axisymmetric velocity perturbations caused by the gap and to highlight the non-axisymmetric features, such as spirals, we subtract the azimuthally averaged velocities (denoted using $<>$) from their native values to create residual maps. We prefer to use the azimuthal averages instead of the Keplerian flow as the background because subtracting the latter results in prominent super- and sub-Keplerian structures in $\vphi$ at the gap edges (Appendix \ref{sec: background}). 

To further compare how planet-induced velocity perturbations vary on 1,000 orbits timescale, we generate difference maps in the right column in each of the three groups in Fig.~\ref{fig: fargo_time}. For example, the radial velocity perturbation difference between 1,000 and 100 orbits \panel{(a3-a1)} is made by subtracting \panel{(a1)} from \panel{(a3)}.

In the radial direction, strong and stable non-axisymmetric velocity structures are present both on and off the spirals (panel group a). At 1,000 orbits (panel (a3)), the former have magnitudes about 170 m/s ($0.06\times\vkepp$, or $0.7\times\cs$, the local sound speed) at $r$ = 85 au at the inner spirals and 180 m/s at 110 au at the outer spirals. 
Both remain relatively unchanged throughout the 1,000 orbits, varying by less than 20\%.

In contrast, $\vrr-<\vrr>$ in the off-spiral regions significantly weaken with time. At 100 orbits (panel (a1)), two structures stand out --- the revolution around L5 inside the gap (magenta ellipse), and the turn of the flow at the two ends of the horseshoe (black arrows). The former has a magnitude of 250 m/s (0.08 $\vkepp$) at 100 orbits, stronger than that of the spirals, before dropping to 200 m/s (0.07 $\vkepp$) at 300 orbits (panel (a2)) and to below 25 m/s at 1,000 orbits (panel (a3)). The horseshoe turn has a magnitude of higher than 300 m/s at 100 orbits, much stronger than that of the spirals, before dropping to 200 m/s around the planet at 1,000 orbits.

In the non-axisymmetric component of the azimuthal velocity, the counterparts of the revolutionary motion around L5 are also prominent at 100 and 300 orbits (panels b1 and b2, magenta ellipses), before damping to below 50 m/s at 1,000 orbits. On the leading and trailing sides of the planet, strong converging flows up to 300 m/s (0.1 $\vkepp$, indicated by the cyan arrows) are visible at 100 orbits, and their magnitudes remain nearly constant throughout 1,000 orbits (panel (b3-b1) and (b3-b2)). Their origin is not entirely clear, as the features are slightly offset from the spiral waves. The non-axisymmetric components at the spirals in the outer disk is below 100 m/s (0.03 $\vkepp$) throughout the simulation,
weaker than that in the inner disk.

In the polar direction, the gas motion with the highest velocity ($>$100 m/s, 0.03 $\vkepp$) is in the infalling flow towards the planet at its vicinity. The area of this flow shrinks and its velocity becomes weaker from 100 to 1,000 orbits (panel (c3-c1)) as the gap get deeper, while its velocity and area remain roughly unchanged between 300 and 1,000 orbits (panel (c3-c2)). 
Turbulent motions at the level of 100 m/s
both inside the gap and at its edges sustained over 1,000 orbits are also prominent, consistent with previous studies \citep[e.g.,][]{dong_observational_2019}. In comparison, the vertical gas motions at the spiral locations are not the dominant signals.

The right column in each of the three groups in Fig.~\ref{fig: fargo_time} shows the differences in the velocity perturbations at two simulation times. Overall, the main differences occur in the non-spiral regions. Comparing \panel{(a3-a1)} with \panel{(a3-a2)}, we find that as time increases, the size of the gas structure at L5 shrinks and the signals at the horseshoe ends become weaker.
Specifically, differences in the radial velocity $\vrd$ at the horseshoe ends between 100 and 1,000 orbits (\panel{(a3-a1)}) reach \(\sim 250\) m/s (0.08 \(\vkepp\)), while the differences in $\vrd$ (and $\vpd$) around L5 between 300 and 1,000 orbits (\panel{(a3-a2)}) reach 150 m/s (0.05 \(\vkepp\)).  
Meanwhile, all three velocities in the spirals vary negligibly between 300 and 1,000 orbits. The strongest variation of $\vrd$ at spirals between 100 and 1,000 orbits appears at $r$ = 120-150 au with a magnitude of 150 m/s (0.05 \(\vkepp\)).

\subsection{Numerical convergence tests} \label{sec:convergence}

\begin{figure*}
\includegraphics[width=0.9\linewidth]{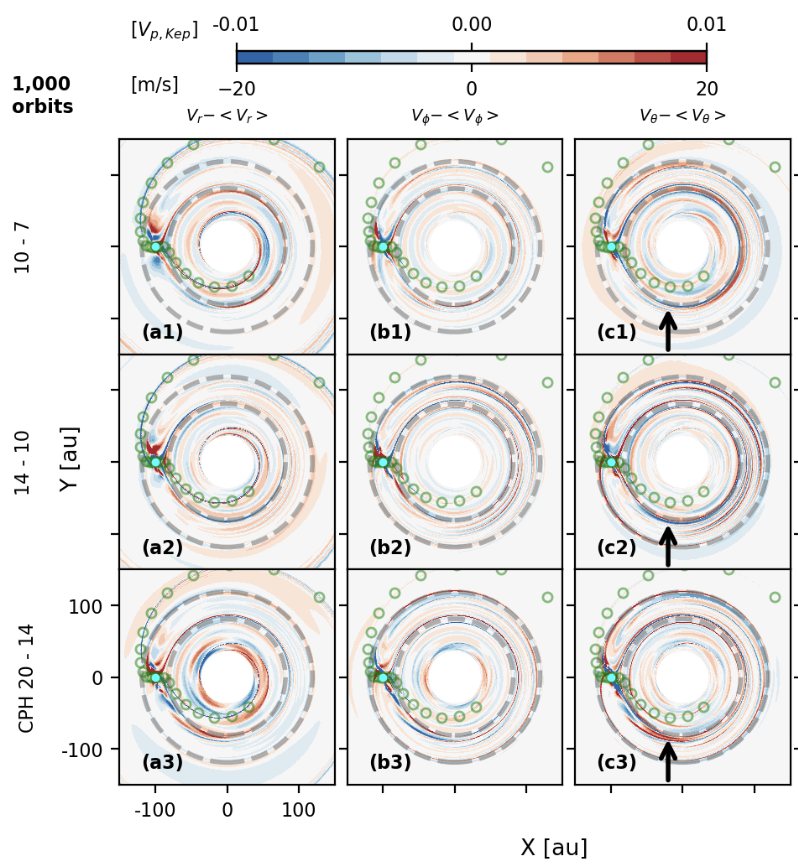} 
\centering
\caption{
Convergence tests with simulations at 4 numerical resolutions: cells per scale height (CPH) around the planet of 7, 10, 14, and 20. The top row shows the differences between two runs with CPH=7 and 10 in non-axisymmetric velocity perturbations at $2h$ away from the midplane at 1,000 orbits for the three velocities. The middle and bottom rows show the differences between the simulations with CPH=10 and 14, and the simulations with CPH=14 and 20, respectively.
The primary inner and outer spiral density waves are marked by open green circles. Grey dashed lines indicate the inner and outer gap edges. The location of the planet is indicated by the cyan marker, and its size is set to 0.4$\rhill$, the expected size of the circumplanetary disk. The colorbar is in a linear scale. See \S\ref{sec:convergence} for discussions.
}
\label{fig:fargo_cph} 
\end{figure*}

We present numerical convergence tests in Fig.~\ref{fig:fargo_cph}. Following the discussions in \S\ref{sec: fargo_time_long}, we focus on 3D non-axisymmetric velocity perturbations at $2h$ at 1,000 orbits. We compare the simulations with different resolutions by showing their differences; for instance, $\panel{(a1)}$ shows the differences in $\vrd$ between two runs with CPH$=10$ and 7. When creating the difference maps, we interpolate the outputs from all lower resolution runs to the highest resolution run, 20 CPH. While the variations are smaller than 20 m/s (0.01 $\vkepp$) in most regions, 
whether any of the three velocity components has converged at our highest resolution (CPH~$=20$) depends on the region in the disk.

\begin{figure*}
\includegraphics[width=\linewidth]{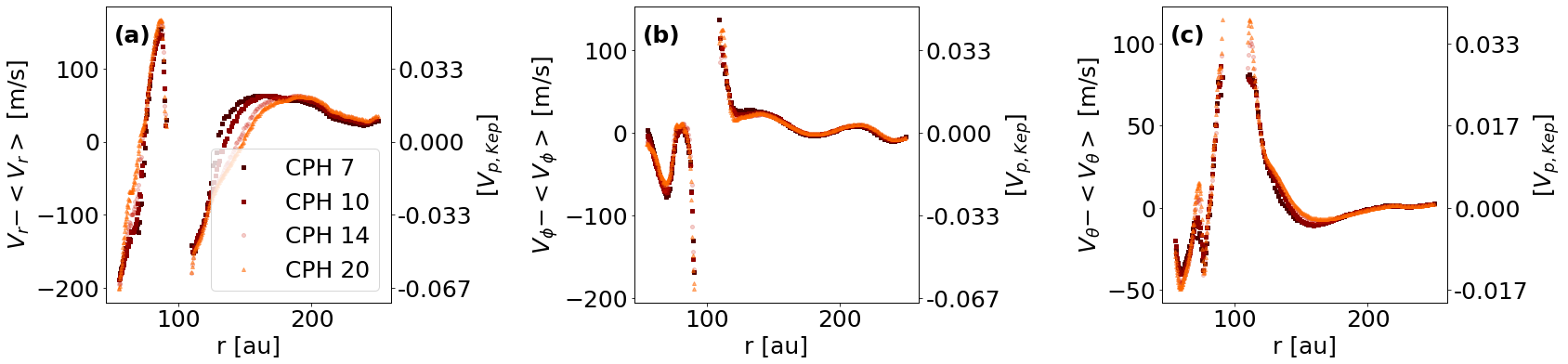} 
\centering
\caption{
The three components in non-axisymmetric velocity perturbations along the primary spiral density waves (traced out by the green circles in Figs.~\ref{fig: sigma_gap} and \ref{fig:fargo_cph}) at $2h$ away from the midplane at 1,000 orbits for 4 simulations with different numerical resolutions: cells per scale height (CPH) of 7, 10, 14, and 20.
We exclude the region within $\rhill$ from the planet at 100 au due to insufficient resolution in the circumplanetary region. 
The velocity field along the spirals is converging with resolution.
See \S\ref{sec:convergence} for further discussions.
}
\label{fig:fargo_cph_sp} 
\end{figure*}

Along the primary spirals (green open circles in Fig. \ref{fig:fargo_cph}), velocity perturbations are converging with resolution. This can be more quantitatively seen in Fig. \ref{fig:fargo_cph_sp}, where we show $\vrd$, $\vpd$, and $\vtd$ along the primary spirals in panels (a), (b), and (c), respectively. 
In contrast, the non-spiral regions have not shown signs of convergence yet at our highest resolution (CPH$=$20). For example, the inner gap edge, indicated by an arrow in Fig. \ref{fig:fargo_cph}, exhibits a larger difference in $\vtd$ between CPH$=$20 and CPH$=$14 (c3), compared with that between CPH$=$14 and CPH$=$10 (c2).

\section{Signatures in synthetic CO observations}
\label{sec:rt}

To explore how the effects of simulation time and resolution manifest in searching for planet-induced kinematic signatures, we post-process the $\fargo$ outputs using $\radmc$ \citep{dullemond_radmc-3d_2012} to generate synthetic observations of \ce{CO} $J$=2-1 line emission. We describe the procedure in \S\ref{sec:rt_setup}, and introduce the results in \S\ref{sec: radmc_times} and \S\ref{sec: radmc_cph}.

\subsection{Synthetic observation generation} \label{sec:rt_setup}

\begin{figure*}
\includegraphics[width=0.7\linewidth]{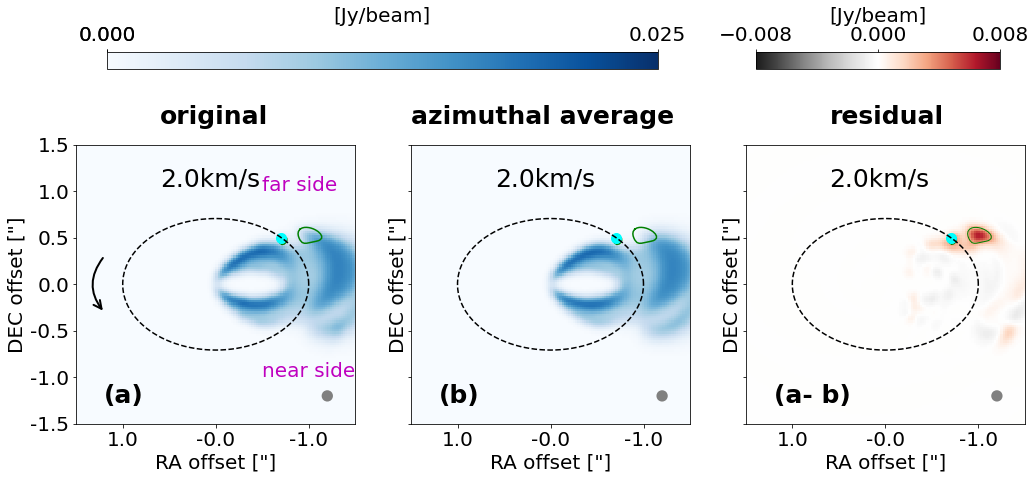} 
\centering
\caption{
Example products from $\radmc$ radiative transfer simulations. Panel (a) is a channel map at $\vch = 2.0 \kms$ directly produced from a FARGO3D simulation output. Panel (b) is a channel map produced from the same FARGO3D simulation, but with density and all three velocities azimuthally averaged. Panel (a-b) is the difference between the two.
Contours in panel (a-b) are at 3$\sigma = $ 3 mJy/beam level (both positive and negative), and they are overplotted in panels (a) and (b). The planet's orbit is indicated by the black dashed ellipse and the arrow indicates the direction of Keplerian rotation. The planet's location is indicated by the cyan marker, and its size is set to 0.4$\rhill$, the expected size of the circumplanetary disk. The disk near side and far side are annotated. 
We put the disk at a distance of 100 pc. 
The synthesized beam is depicted in the bottom right of each panel. The colorbars are in linear scale. 
} 
\label{fig: radmc_intro}
\end{figure*}

We produce synthetic CO observations by sending the $\fargo$ output density and velocity structures into $\radmc$ to generate channel maps at our specified spectral resolution. They are then convolved by a point spread function to achieve the desired angular resolution. An example channel map is shown in Fig.~\ref{fig: radmc_intro} (panel a).

To set up $\radmc$, we assume that the entire disk extends from 1 to 275 au, with a disk mass of 1.7$\times 10^{-3} \Msun$. For the extrapolation from 36.5 au, the inner boundary of the hydro simulations, to 1 au, we assume $\Sigma(r) \propto r^{-1}$.
For the stellar parameters, we adopt $\Mstar = 1\Msun$, $\Rstar = 1.7\Rsun$, and $\Tstar = 4,700$K, suitable for a pre-main-sequence star. 
Stellar radiation serves as the sole heating source. $\nphot = 10^8$ photon packages are used. 
We assume silicate dust particles with an intrinsic density of 3.71 g/cm$^{3}$. We include grains with sizes ranging from 0.1 to $10\mu m$ in the calculation. Due to the effective coupling of such small dust grains with the gas, the dust and gas are assumed to be well-mixed. We also assume the grain size distribution follows a power law with an index of -3.5 and a maximum grain size of 1mm (grains with sizes between $10\mu m$ and 1 mm are not included in the simulations), and the total dust to gas mass ratio is 1$\%$. Therefore, the dust mass within 0.1 to $10\mu m$ is about 10\% of the total dust mass, or 0.1$\%$ of the total disk mass. We calculate the corresponding dust opacity using the optool package \citep{dominik_optool_2021}.

To make synthetic channel maps, we assume the gas temperature is the same as the dust temperature from radiative transfer. We assume a CO to H$_2$ ratio of $10^{-4}$, and a viewing angle of disk inclination $=45\degree$ (similar to that of disk HD 163296; \citealt{pinte_kinematic_2018}), position angle (PA) $=90\degree$, and the southern side being the near side. The images are convolved with a $0.1 \arcsec$ Gaussian beam, typical in observations of kinematic signatures \cite[][Table A2]{speedie_testing_2022}.
The synthetic cubes have a channel width of 0.2 km/s.

Negative (positive) velocities denote gas moving towards (away from) the observer along the line of sight (LOS). A planet is placed at a position angle of $\phip = 315 \degree$ in the disk frame, in between the disk major and minor axes. The planet orbits the star in the counter-clockwise direction at $\rp=$ 100 au with a LOS velocity $\vlosp =1.5$ $\kms$. 

To highlight and quantify the non-axisymmetric features in gas emission caused by the planet, we create residual channel maps, illustrated in Fig.~\ref{fig: radmc_intro} (panel c). Similar to the residual velocity maps (\S\ref{sec: fargo_time_long}; Fig.~\ref{fig: fargo_time}), residual channel maps are also produced by subtracting the azimuthally averaged background (we azimuthally average the hydro quantities before conducting the radiative transfer) instead of the Keplerian background from the original channel maps. In real observations, the azimuthal average background may be found by fitting the observational data with models using public tools such as eddy \citep{teague_eddy_2019}.
Green contours in the residual panel mark $3\sigma$ regions that are larger than a beam size, and are overlaid in the other panels. We adopt $\sigma=$ 1 mJy/beam, corresponding to 10 hours of integration time with our observing parameters\footnote{\url{https://almascience.eso.org/proposing/sensitivity-calculator}}. In the residual channel map at a specific velocity $\vch$, a positive region (red) has more emission at this velocity in the planet-perturbed disk than that in the azimuthally-averaged disk, and vise versa.

\begin{figure*}
\includegraphics[width=\linewidth]{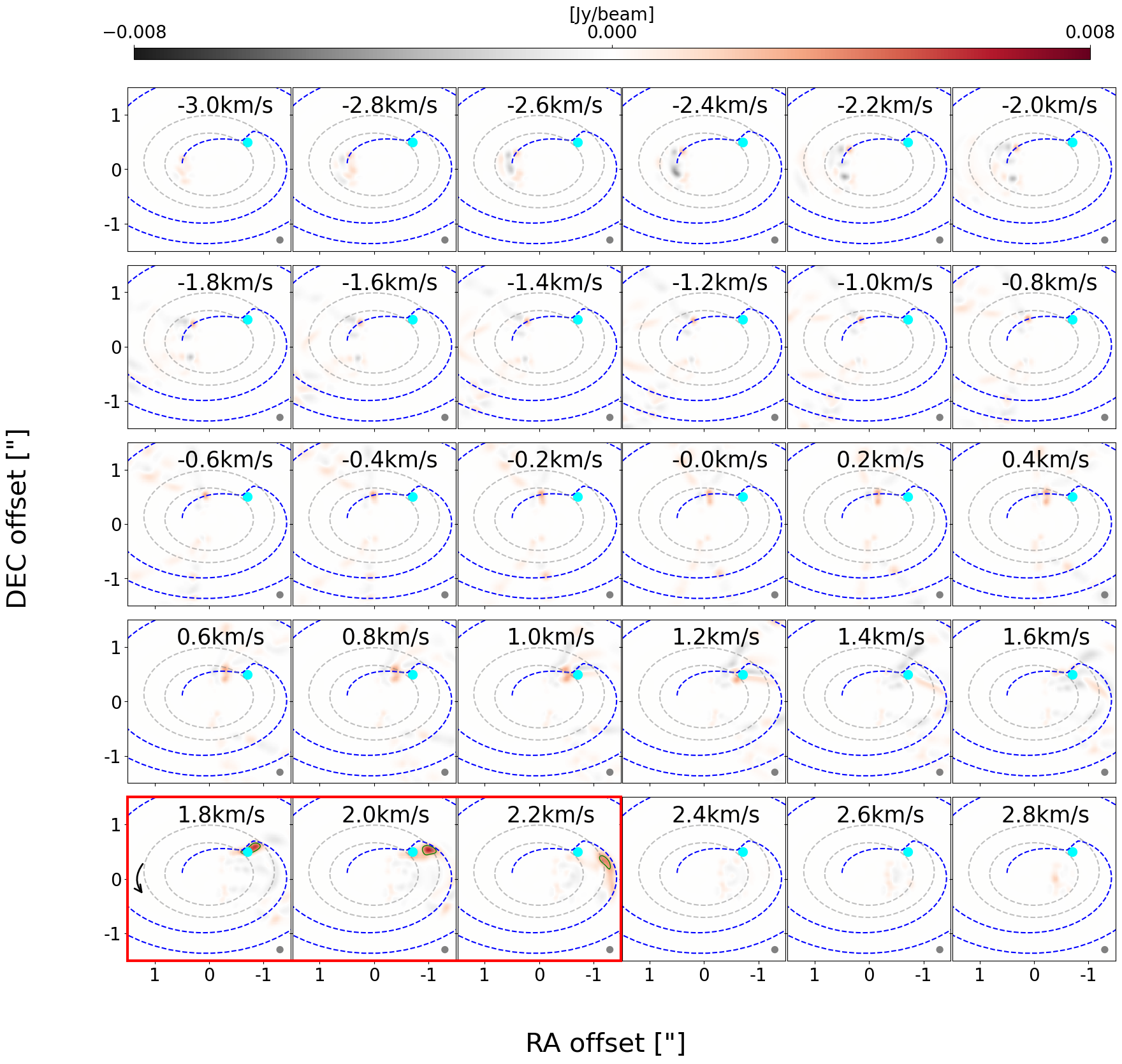} 
\centering
\caption{
Successive residual channel maps of the same kind as panel (a-b) in Fig.~\ref{fig: radmc_intro} of the simulation with CPH$=20$ at 1,000 orbits. Signals in the maps highlight non-axisymmetric velocity perturbations induced by the planet. Potentially detectable residuals are highlighted by $3\sigma$ contours (green). We only mark $3\sigma$ signals with a size bigger than a beam. Such signals are most prominently present in channels at $\vch = 1.8$ to $2.2$ $\kms$ (highlighted with a red frame). Grey and blue dashed lines denote gap edges and the primary spirals, respectively. 
The location of the planet is indicated by the cyan marker. The disk rotation direction is marked in the lower left panel. The colorbar is in a linear scale.
See \S\ref{sec:rt_setup} for further discussions.
} 
\label{fig: chan_resi}
\end{figure*}

We present successive residual channel maps of the same type as Fig.~\ref{fig: radmc_intro}c from the simulation with CPH$=$20 at 1,000 orbits in Fig.~\ref{fig: chan_resi}. The corresponding original channel maps are shown in Appendix \ref{sec: chan}. Potentially detectable ($>3 \sigma$) non-axisymmetric signals are present in some channels (3$\sigma$ contours marked in green), specifically in the velocity range 
$\vch$ = 1.8 to 2.2 $\kms$ (the panels enclosed in a red frame). 
The $3\sigma$ features at $\vch$ = 1.8 and 2.0 $\kms$ coincide with the planet. The $3\sigma$ feature at 2.2 km/s is extended, and slightly offset from the outer primary spiral and partly overlapping with the outer gap edge.

\subsection{Effect of simulation time} \label{sec: radmc_times}

\begin{figure*}
\includegraphics[width=0.9\linewidth]{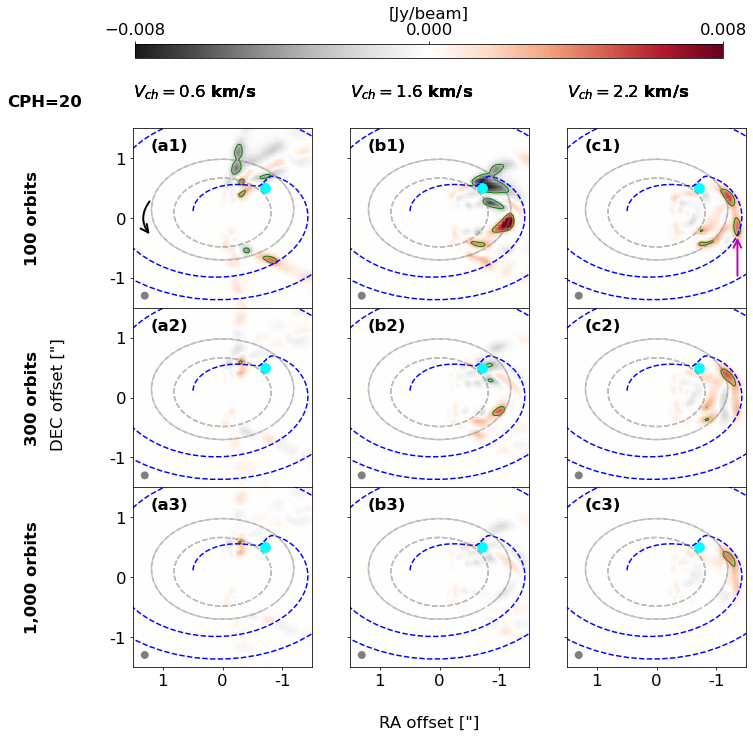} 
\centering
\caption{
Residual channel maps (the same kind as panel (a-b) in Fig.~\ref{fig: radmc_intro}) of the simulation with CPH$=20$ at $\vch=$ 0.6 (the (a) panels), 1.6 (b), and 2.2 $\kms$ (c). Results at 100, 300, and 1,000 orbits are shown from top to bottom. 
Planet-induced non-axisymmetric kinematic signatures evolve with time. 
The planet is positioned at the position angle $\phip=315 \degree$ in the disk frame. 
Residual emissions stronger than 3$\sigma = \pm$ 3.0 mJy/beam are marked with green contours. In each panel, the blue dashed curve represents the primary spirals at $2h$ away from the midplane, and grey lines mark the inner and outer gap edge. The beam size is displayed at the bottom left. The disk rotation direction is marked in panel (a1). The colorbar is in a linear scale. See \S\ref{sec: radmc_times} for discussions.
}
\label{fig: radmc_time}
\end{figure*}

Fig.~\ref{fig: radmc_time} illustrates how planet-induced non-axisymmetric kinematic signatures, manifesting as features in residual channel maps, evolve with time over 1,000 orbits. We focus on 3 channels, $\vch=$ 0.6, 1.6, and 2.2 $\kms$.

Overall, regions with $3\sigma$ residuals shrink with time.
At $\vch=0.6$ km/s (the (a) panels), a channel far away from $\vlosp$ = 1.5 km/s,  $3\sigma$ signals
appear at the outer gap edge at PA $\sim$ 0 and near the outer primary spiral at 100 orbits (panel a1). However, such signals disappear at 300 and 1,000 orbits. At $\vch=1.6$ km/s,
$3\sigma$ residuals at L5 at PA$\sim$ 270 $\degree$ are present at both 100 (b1) and 300 (b2) orbits, so do residuals at horseshoe turns at 100 orbits. In contrast, there is no $3\sigma$ residual signal at 1,000 orbits.

In addition, the morphology of the robust (long-lasting) $3\sigma$ residuals also evolves with time. At $\vch=2.2$ km/s (the (c) panels), although the features crossing the outer gap edge are visible at all three epochs, their sizes slightly decrease with time. At 100 orbits (c1), the $3\sigma$ signal along the outer primary spiral splits into two regions. One of the two regions in panel (c1), indicated by a magenta arrow, drops below $3\sigma$ at 300 and 1,000 orbits (c2 and c3).

\subsection{Effects of numerical resolution}
\label{sec: radmc_cph}

\begin{figure*}
\includegraphics[width=\linewidth]{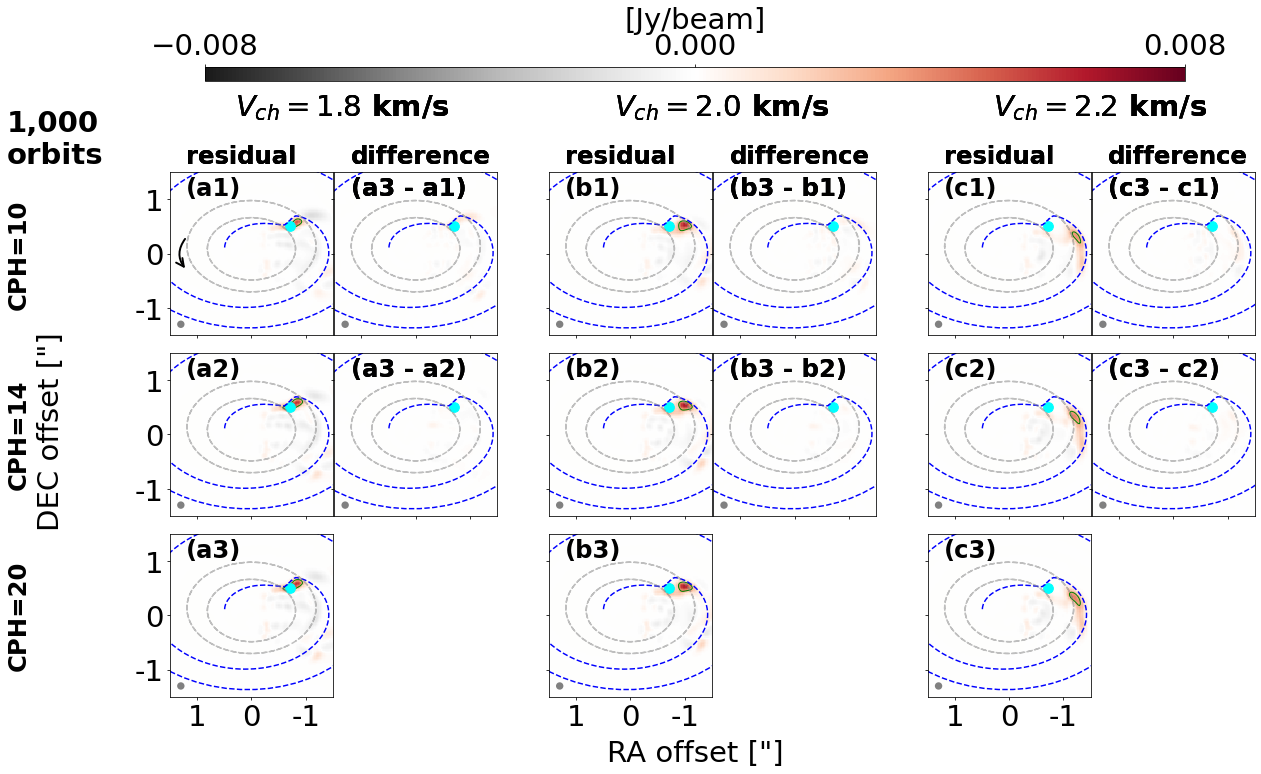} 
\centering
\caption{
Resolution convergence test of synthetic channel maps. For each $\vch$, the left and right columns present residuals (similar to Fig.~\ref{fig: radmc_intro}(a-b)) and differences, respectively. Difference channel maps are obtained by subtracting residual channel maps at different resolutions. Residual or difference emission stronger than 3$\sigma = \pm$ 3.0 mJy/beam are marked with green contours. The disk rotation direction is marked in the top left panel. The colorbar is in a linear scale. See \S\ref{sec: radmc_cph} for discussions.
} 
\label{fig: radmc_cph}
\end{figure*}

Fig. \ref{fig: radmc_cph} shows the residual and difference channel maps at $\vch= 1.8 - 2.2$ km/s at 1,000 orbits for simulations with resolutions of CPH$=$10, 14 and 20. Difference channel maps are obtained by subtracting residual channel maps at different resolutions. The results exhibit good convergence. No 3$\sigma$ features are found in the difference maps, both at the three velocities and at all other velocities (not shown here). 
Therefore, given the chosen sensitivity, angular resolution, and channel width, the differences due to different numerical resolutions ranging from 10 to 20 CPH cannot be discerned in synthetic observations.

\section{conclusions} 
\label{sec: conclusion}

We use grid-based hydrodynamic and radiative transfer simulations to investigate how planet-induced kinematic signatures depend on simulation time and numerical resolution in both velocity space and synthetic CO line emission. We focus on non-axisymmetric signals that are more easily localized in observations. We choose a planet with 2.5 $\Mj$ (5 disk thermal masses) at 100 au, typical for planets detected via local kinematic signatures (e.g., ``kinks''; \citealt{pinte_kinematic_2018, pinte_kinematic_2019, pinte_nine_2020}). We focus on signatures at  2 disk scale heights, close to the CO emission surface. We propose to identify and quantify planet-induced kinematic signals in residual channel maps, the difference between the original channel map and the one produced from the same disk-planet model with density and velocities azimuthally averaged (panel (a-b) in Fig.~\ref{fig: radmc_intro}). Our main findings are: 
\begin{enumerate}
    \item Simulations of short timescales, e.g., 100 orbits, are insufficient for establishing steady planet-induced velocity perturbations in grid-based simulations with viscosity $\alpha=10^{-3}$. We find strong velocity structures at non-spiral regions, including the Lagrange points, horseshoe, and gap edges, at 100 orbits (Figs.~\ref{fig: hd_100ob}). Their strengths can be comparable to, or even bigger than, the more robust and steady signatures along the spirals. In addition, such signatures vary significantly on dynamical timescale (e.g. exceeding 250 m/s or 0.08 $\vkepp$ as shown in Fig.~\ref{fig: vr_times}). They are potentially detectable in ALMA CO observations with 10 hours of integrations (Fig.~ \ref{fig: radmc_time}). 
    However, these features are damped over 1,000 orbits. 

    \item A sufficiently long simulation time, such as 1,000 orbits, can establish a steady velocity field. Based on results at 1,000 orbits shown in Fig.~\ref{fig: fargo_time}, the strongest velocity deviations from the azimuthal average background can reach up to $\pm300$ m/s (0.1 $\vkepp$) in the azimuthal direction near the planet and slightly offset from the density spiral. 
    
    \item Robust non-axisymmetric velocity perturbations at 1,000 orbits can be present in several velocity channels in residual channel maps. In our setup with median disk inclination and planet position angle in between the disk major and minor axes  (similar to the case in \citet{pinte_kinematic_2018}), such perturbation signal appears around the line-of-sight velocity of the planet. They are potentially detectable in ALMA programs with reasonable parameters (an angular resolution of 0.1$''$, a channel width of 200 m/s, and a sensitivity of 1 mJy/beam; Fig.~\ref{fig: chan_resi}).
    
    \item At 1,000 orbits, planet-induced velocity perturbation along the spirals converges in hydro simulations at a numerical resolution of 20 cells per scale height (CPH). The results obtained from the runs with CPH$=14$ and CPH$=20$ differ by less than 10 m/s in all three velocities (Figs.~\ref{fig:fargo_cph} and \ref{fig:fargo_cph_sp}). However, convergence is not achieved in off-spiral regions even at our highest resolution (CPH$=20$). 
    Nevertheless, the effect
    may not be detectable in ALMA observations with our observing setup (Fig.~\ref{fig: radmc_cph}).
\end{enumerate}

\section*{Acknowledgments}

We are thankful to the referee for the constructive report. We thank Jeremy Smallwood, Dhruv Muley for discussion. K.C. acknowledges support by UCL Dean’s Prize and China Scholarship Council.
R.D. acknowledges financial support provided by the Natural Sciences and Engineering Research Council of Canada through a Discovery Grant, as well as the Alfred P. Sloan Foundation through a Sloan Research Fellowship. This research was enabled in part by support provided by the Digital Research Alliance of Canada \url{alliance.can.ca}.


\bibliography{zotero}{}
\bibliographystyle{aasjournal}

\appendix

\section{$\tau=1$ surface}
\label{sec: tau1}

\begin{figure*}
\includegraphics[width=0.5\linewidth]{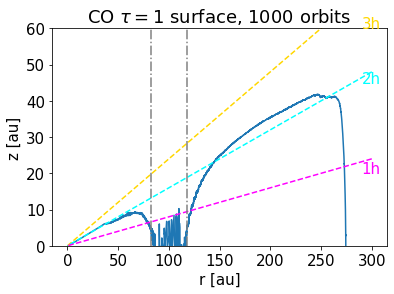} 
\centering
\caption{Azimuthally averaged $\tau=1$ surface of synthetic \ce{CO} $J=2-1$ emission (blue line) of the simulation with CPH$=20$ at 1,000 orbits. Dashed lines in magenta, cyan, and yellow represent 1, 2, and 3 scale heights, respectively. The grey dashed-dotted lines mark the inner and outer gap edge. The planet is at $r=100$ au.
}
\label{fig: tau1} 
\end{figure*}

In Fig.~\ref{fig: tau1}, we show the azimuthal average $\tau=1$ surface of CO $J=2-1$ emission of the simulation with CPH$=20$ at 1,000 orbits. The $\tau=1$ surface is determined under the assumption that the disk inclination is 0. Except for the deep gap region (marked by grey dashed-dotted lines), with the $\tau=1$ surface is roughly located at 2 scale heights.

\section{Local variabilities in $\vphi - \vkep$ and $\vtheta$ on orbital timescale}
\label{sec: short_term_others}
\begin{figure*}
\includegraphics[width=\linewidth]{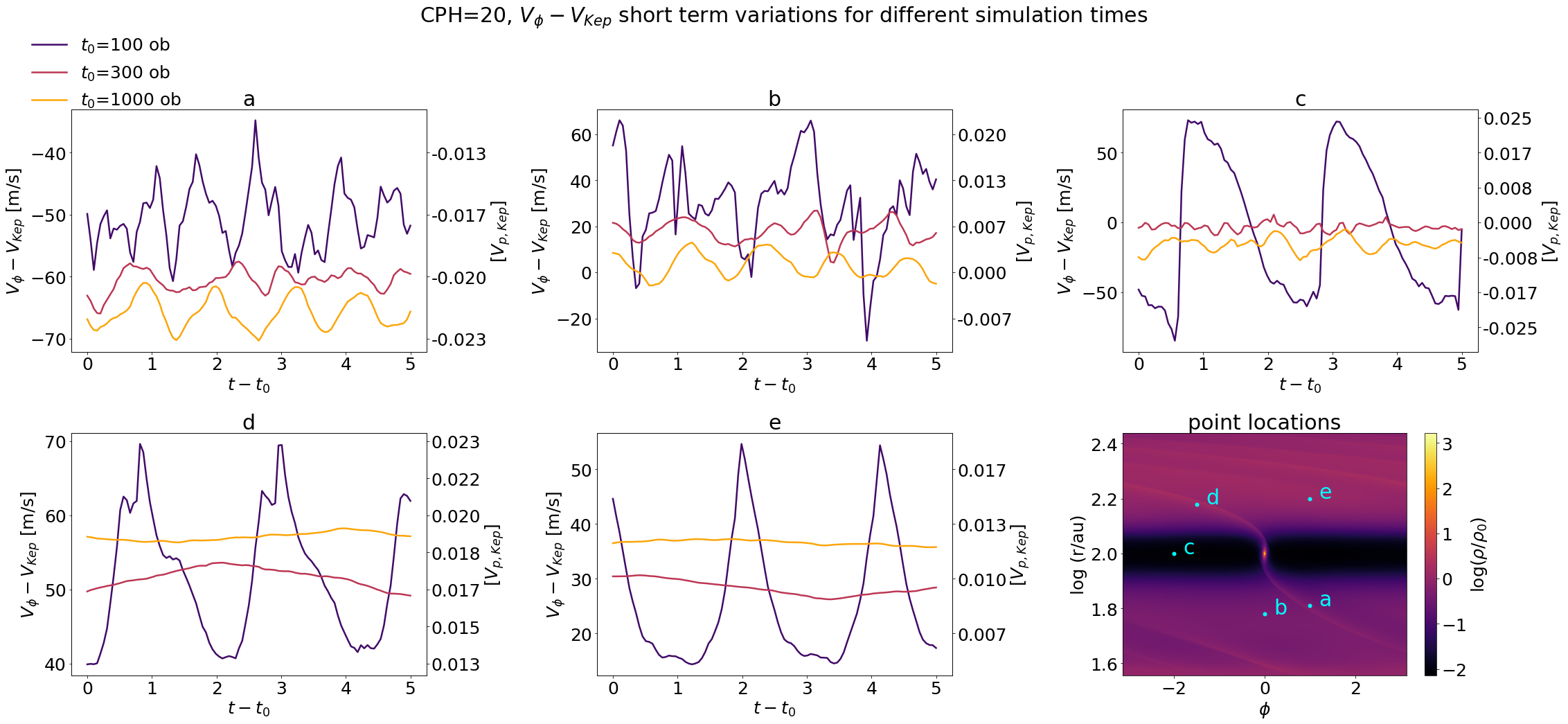} 
\centering
\caption{Similar to Fig.~\ref{fig: vr_times} but for $\vphi - \vkep$.
}
\label{fig: dvphi_times} 
\end{figure*}

\begin{figure*}
\includegraphics[width=\linewidth]{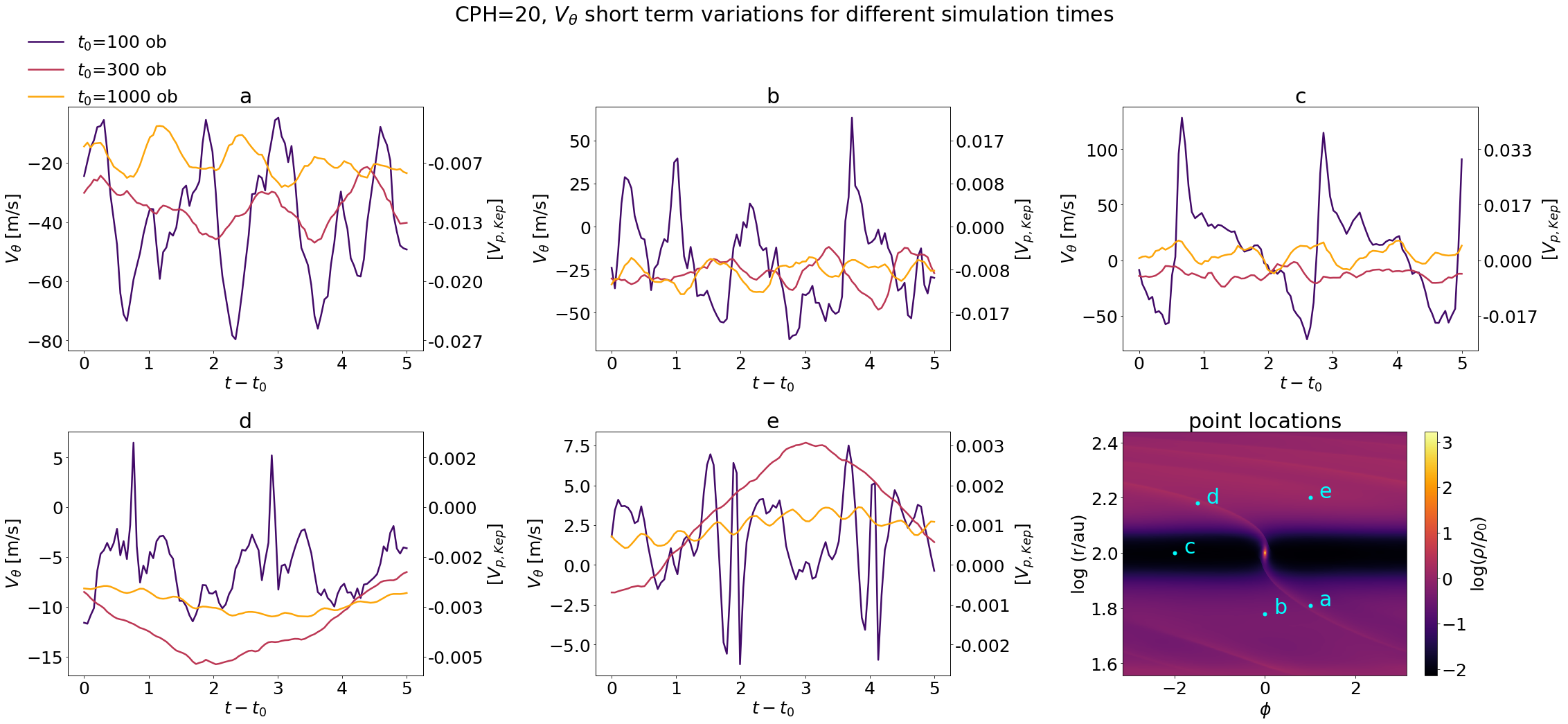} 
\centering
\caption{Similar to Fig.~\ref{fig: vr_times} but for $\vtheta$.
}
\label{fig: vtheta_times} 
\end{figure*}

Fig.~\ref{fig: dvphi_times} and ~\ref{fig: vtheta_times} show the short term variabilities within 5 orbits of $\vphi - \vkep$ and $\vtheta$ at $2h$ at 100, 300. and 1,000 orbits at the same 5 representative locations as in \S\ref{sec: fargo_time_short} and Fig. \ref{fig: vr_times}. Similar to $\vrr$ (\S\ref{sec: fargo_time_short}), both $\vphi - \vkep$ and $\vtheta$ display significant short term variations at 100 orbits (e.g. point c, they vary more than 0.05 $\vkepp$) while the variability drops at 300 and further at 1,000 orbits (5-orbit temporal STDs less than 20 m/s for all 5 points).

\section{Azimuthal average background versus Keplerian background}
\label{sec: background}

\begin{figure*}
\includegraphics[width=0.7\linewidth]{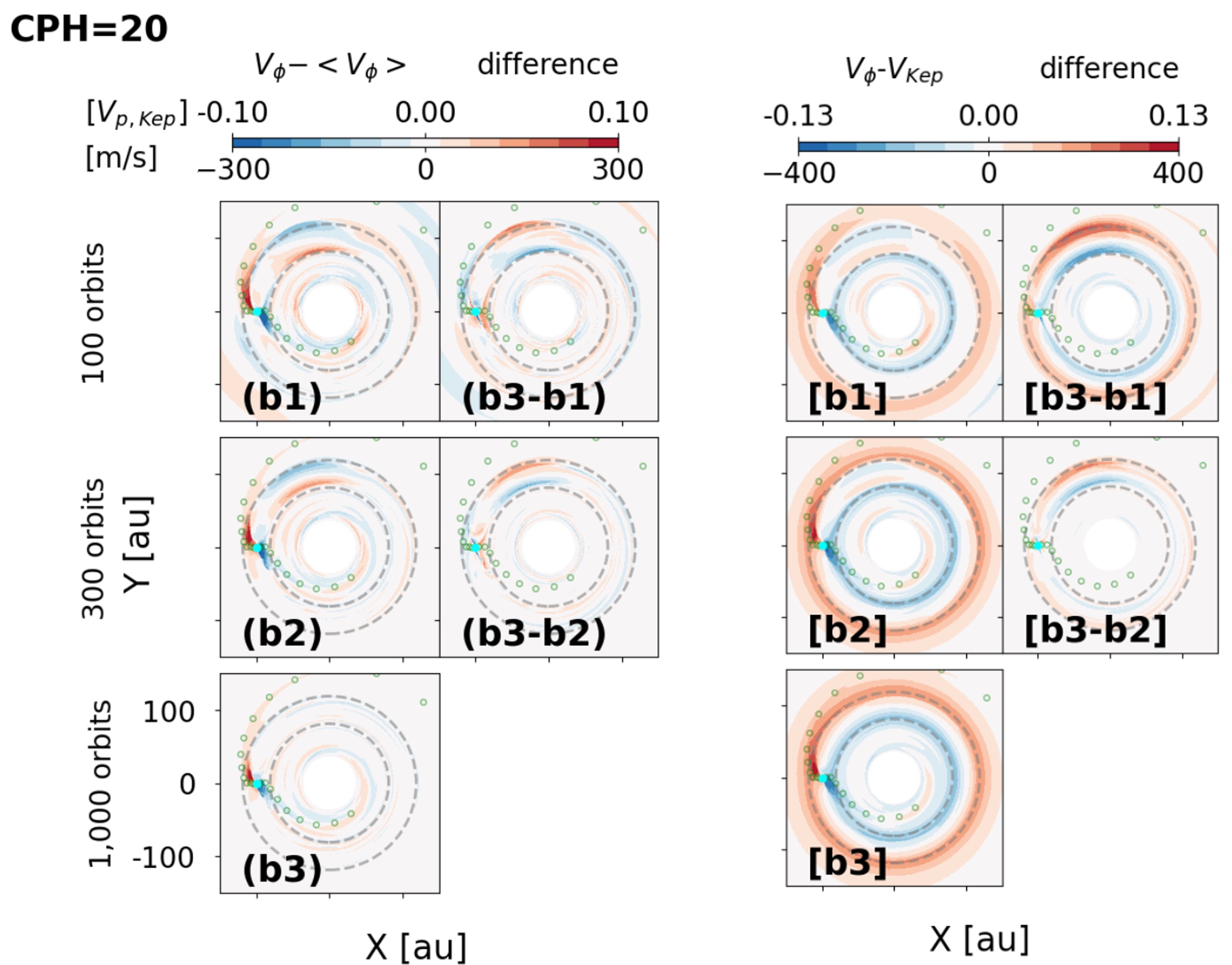} 
\centering
\caption{
Comparison of $\vphi$ perturbations obtained by subtracting azimuthal average background (left) and Keplerian background (right) from original $\fargo$ simulation results. Note that the colorbars are in linear scale and the ranges differ between these two groups of panels.
} 
\label{fig: fargo_bg}
\end{figure*}

Fig. \ref{fig: fargo_bg} compares the $\vphi$ perturbations when subtracting the azimuthally averaged (left) and $\kep$ background (right) from original $\fargo$ results. Note that we do not show radial and colatitude velocity perturbations here since they are similar in both cases. With the $\kep$ background subtracted, stronger $\vphi - \vkep$ signals emerge as simulation time increases, attributed to the steeper pressure gradients at the gap edges at later time. Both inner and outer gap edges (gray dashed lines) exhibit perturbations with an amplitude of $\sim$250 m/s (0.08 $\vkepp$) at 1,000 orbits, complicating the quantification of the contribution from spirals. Given our focus on non-axisymmetric kinematic signals, subtracting the azimuthally averaged background is more suitable.

\section{Channel maps}
\label{sec: chan}

\begin{figure*}
\includegraphics[width=\linewidth]{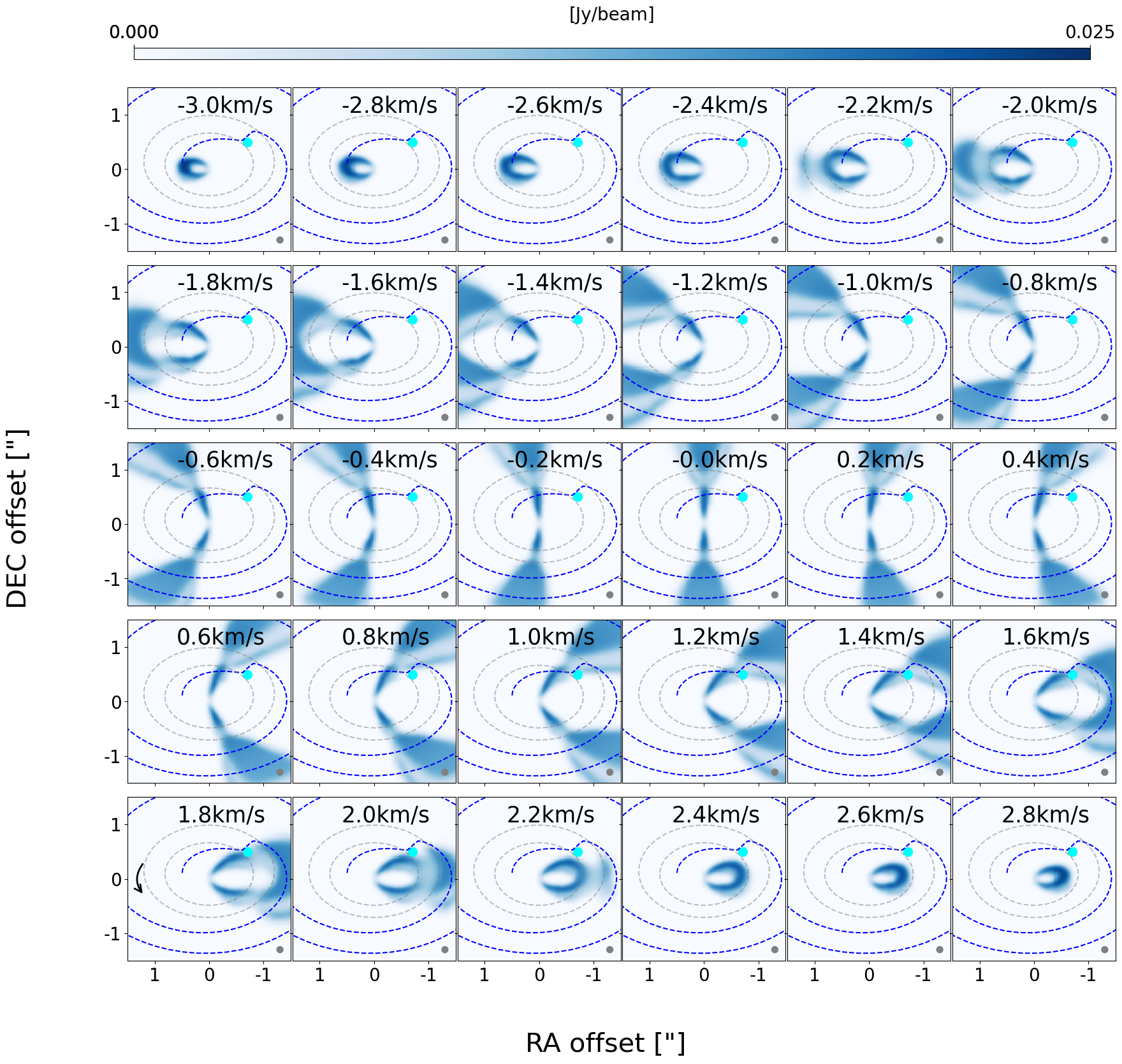} 
\centering
\caption{Successive channel maps of our disk model at 1,000 orbit and with CPH=20. Grey and blue dashed lines denote gap edges and the primary spirals, respectively. 
The location of the planet is indicated by the cyan marker. The disk rotation direction is marked in the lower left panel. The colorbar is in a linear scale.
} 
\label{fig: radmc_chan}
\end{figure*}

Successive channel maps (counterparts of Fig.~\ref{fig: chan_resi}) of our disk model at 1,000 orbits and with CPH=20 are shown in Fig.~\ref{fig: radmc_chan}.

\end{document}